\documentclass[twocolumn,compsoc,10pt,journal]{IEEEtran}
\usepackage[T1]{fontenc}
\usepackage{array}
\usepackage{float}
\usepackage{calc}
\usepackage{units}
\usepackage{url}
\usepackage{amsthm}
\usepackage{amsmath}
\usepackage{graphicx}
\usepackage[unicode=true,
 bookmarks=true,bookmarksnumbered=true,bookmarksopen=true,bookmarksopenlevel=1,
 breaklinks=false,pdfborder={0 0 0},backref=false,colorlinks=false]
 {hyperref}
\hypersetup{pdftitle={Your Title},
 pdfauthor={Your Name},
 pdfpagelayout=OneColumn, pdfnewwindow=true, pdfstartview=XYZ, plainpages=false}
\usepackage{breakurl}

\makeatletter

\providecommand{\tabularnewline}{\\}
\newcommand{\lyxdot}{.}

\floatstyle{ruled}
\newfloat{algorithm}{tbp}{loa}
\providecommand{\algorithmname}{Algorithm}
\floatname{algorithm}{\protect\algorithmname}

 \let\oldforeign@language\foreign@language
 \DeclareRobustCommand{\foreign@language}[1]{%
   \lowercase{\oldforeign@language{#1}}}
\theoremstyle{plain}
\newtheorem{thm}{\protect\theoremname}

\usepackage[caption=false,font=normalsize,labelfont=sf,textfont=sf]{subfig}

\makeatother

\providecommand{\theoremname}{Theorem}

\begin{document}

\title{Limited Random Walk Algorithm for Big Graph Data Clustering}

\author{Honglei~Zhang,~\IEEEmembership{Member,~IEEE,} Jenni~Raitoharju,~\IEEEmembership{Member,~IEEE,}
Serkan~Kiranyaz,~\IEEEmembership{Senior Member,~IEEE,} and~Moncef~Gabbouj,~\IEEEmembership{Fellow,~IEEE}\IEEEcompsocitemizethanks{\IEEEcompsocthanksitem Honglei~Zhang, Jenni~Raitoharju, and Moncef~Gabbouj
are with the Department of Signal Processing, Tampere University of
Technology, Tampere, Finland\protect \\
E-mail: \href{http://{honglei.zhang, jenni.raitoharju, moncef.gabbouj}@tut.fi}{{honglei.zhang, jenni.raitoharju, moncef.gabbouj}@tut.fi}.

\IEEEcompsocthanksitem Serkan~Kiranyaz is with the Electrical Engineering
Department, College of Engineering, Qatar University, Qatar\protect \\
E-mail: \href{mailto:mkiranyaz@qu.edu.qa}{mkiranyaz@qu.edu.qa}

}\thanks{}}

\markboth{}{Honglei Zhang \MakeLowercase{\textit{et al.}}: Mr}

\IEEEtitleabstractindextext{
\begin{abstract}
Graph clustering is an important technique to understand the relationships
between the vertices in a big graph. In this paper, we propose a novel
random-walk-based graph clustering method. The proposed method restricts
the reach of the walking agent using an inflation function and a normalization
function. We analyze the behavior of the limited random walk procedure
and propose a novel algorithm for both global and local graph clustering
problems. Previous random-walk-based algorithms depend on the chosen
fitness function to find the clusters around a seed vertex. The proposed
algorithm tackles the problem in an entirely different manner. We
use the limited random walk procedure to find attracting vertices
in a graph and use them as features to cluster the vertices. According
to the experimental results on the simulated graph data and the real-world
big graph data, the proposed method is superior to the state-of-the-art
methods in solving graph clustering problems. Since the proposed method
uses the embarrassingly parallel paradigm, it can be efficiently implemented
and embedded in any parallel computing environment such as a MapReduce
framework. Given enough computing resources, we are capable of clustering
graphs with millions of vertices and hundreds millions of edges in
a reasonable time. \end{abstract}


}

\maketitle

\IEEEdisplaynontitleabstractindextext{}

\IEEEpeerreviewmaketitle{}

\section{Introduction}

\IEEEPARstart{G}{raph} data are important data types in many scientific
areas, such as social network analysis, bioinformatics, and computer
and information network analysis. Big graphs are normally heterogeneous.
They have such non-uniform structures that edges between vertices
in a group are much denser than edges connecting vertices in different
groups. Graph clustering (also named as ``community detection''
in the literature) algorithms aim to reveal the heterogeneity and
find the underlying relations between vertices. This technique is
critical for understanding the properties, predicting dynamic behavior
and improving visualization of big graph data. 

Graph clustering is a computationally challenging and difficult task,
especially for big graph data. Many algorithms have been proposed
over the last decades \cite{schaeffer_graph_2007}. The criteria-based
approaches try to optimize clustering fitness functions using different
optimization techniques. Newman defined a modularity measurement based
on the probability of the link between any two vertices. He applied
a greedy search method to minimize this modularity fitness function
in order to partition a graph into clusters \cite{newman_fast_2004}.
Blondel and Clauset used the same fitness function but combined it
with other optimization techniques \cite{blondel_fast_2008,clauset_finding_2004,waltman_smart_2013}.
Spielman and Teng opted the graph conductance measurement as the fitness
function \cite{spielman_local_2008}. Other than criteria-based methods,
spectral analysis has also been widely adapted in this area \cite{qiu_graph_2006,spielman_spectral_2011}.
Random-walk-based methods tackle the problem from a different angle
\cite{dongen_graph_2000,macropol_rrw:_2009}. These methods use the
Markov chain model to analyze the graph. Each vertex represents a
state and the links indicate transitions between the states. The probability
values that are distributed among the states (vertices) reveal the
graph structure. 

In recent years, the size of graph data has grown dramatically. Furthermore,
most graphs are highly dynamic. It is very challenging or even intractable
to partition the whole graph in real-time. Very often, people are
only interested in finding the cluster that contains a seed vertex.
This problem is called local clustering problem \cite{spielman_local_2008,chung_local_2013,macko_local_2013}.
For example, from an end user's perspective, finding the closely connected
friends around him or her is more important than revealing the global
user clusters of a large social network. It is unnecessary to explore
the whole graph structure for this problem. Recently, random walk
methods have gained great attention on this local graph clustering
problem, since a walk started from the seed vertex is more likely
to stay in the cluster where the seed vertex belongs. Comparing to
the criteria-based methods, the random-walk-based methods are capable
of extracting local information from a big graph without the knowledge
of the whole graph data. In \cite{andersen_using_2007,buhler_constrained_2013,zhu_local_2013},
a random walk is first applied to find important vertices around the
seed vertex. Then a sweep stage is involved to select the vertices
that minimize the conductance of the candidate clusters. 

The accuracy of any criteria-based clustering method (or those combined
with the random walk procedures) is greatly affected by the chosen
clustering fitness function. Furthermore, most local clustering algorithms
use the criteria that are more suitable for the global graph clustering
problem. These choices greatly degrade the performance of these algorithms
when the graph is big and highly uneven. Also the majority of the
graph clustering algorithms are designed in sequential computing paradigm.
Therefore, they do not take advantage of modern high-performance computing
systems. 

In this paper, we propose a novel random-walk-based graph clustering
algorithm---the Limited Random Walk (LRW) algorithm. First of all,
the LRW algorithm does not rely on any clustering fitness function.
Furthermore, the proposed method can efficiently tackle the computational
complexity using a parallel programming paradigm. Finally, as a unique
property among many graph clustering methods, the LRW can be adapted
to both global and local graph clustering in an efficient way. 

The rest of the paper is organized as follows: basics of the graph
structure and the random walk procedure are explained in Section 2;
the LRW procedure is explained and analyzed in Section 3; the proposed
LRW algorithm for the global and local graph clustering problems are
introduced in Section 4; an extensive set of experiments on the simulated
and real graph data, along with both numerical and visual evaluations
are given in Section 5; finally, the conclusions and future work are
in Section 6.

\section{Basic Definitions and the Random Walk Procedure }

Let $G(V,E)$ denote a graph of $n$ vertices and $m$ edges, where
$V=\left\{ v_{i}|i=1,\ldots n\right\} $ is the set of vertices and
$E=\left\{ e_{i}|i=1,\ldots m\right\} $ is the set of edges. Let
$A\in\mathcal{R}^{n\times n}$ be the adjacency matrix of the graph
$G$ and $A_{ij}$ are the elements in the matrix $A$. Let $D\in\mathcal{R}^{n\times n}$
be the degree matrix, which is a diagonal matrix whose elements on
the diagonal are the degrees of each vertex. In this paper, we assume
the graph is undirected, unweighted and does not contain self-loops. 

Clustering phenomenon is very common in big graph data. A cluster
in a graph is a vertex set where the density of the edges inside the
cluster is much higher than the density of edges that link the inside
vertices and the outside vertices. 

Random walk on a graph is a simple stochastic procedure. At the initial
state, we let an agent stay on a chosen vertex (seed vertex). At each
step, the agent randomly picks a neighboring vertex and moves to it.
We repeat this movement and study the probability that the agent lands
on each vertex. 

Let $x_{i}^{(t)}$ denote the probability that the agent is on vertex
$v_{i}$ after step $t$, where $i=1,2,\ldots n$. $x_{i}^{(0)}$
is the probability of the initial state. Let $s$ be the seed vertex.
We have $x_{s}^{(0)}=1$, and $x_{i}^{(0)}=0$ for $i\ne s$. Let
$x^{(t)}=\left[x_{1}^{(t)},x_{2}^{(t)},\ldots,x_{n}^{(t)}\right]^{T}$
be the probability vector, where the superscript $T$ denotes the
transpose of a matrix or a vector. By the definition of the probability,
it is easy to see that $\sum_{i=1}^{n}x_{i}^{(t)}=1$ or $\left\Vert x^{(t)}\right\Vert _{1}=1$. 

The random walk procedure is equivalent to a discrete-time stationary
Markov chain process. Each vertex is corresponding to a state in the
Markov chain and each edge indicates a possible transition between
the two states. The Markov transition matrix $P$ can be obtained
by normalizing the adjacency matrix to have each column sum up to
1, e.g. 
\begin{equation}
P_{ij}=\frac{A_{ij}}{\sum_{k=1}^{n}A_{kj}}\label{eq:transition_matrix_element}
\end{equation}
 or 
\begin{equation}
P=AD^{-1}.\label{eq:classifical_transition_matrix}
\end{equation}
 Other forms of the transition matrix $P$ can also be used, for example
the lazy random walk transition matrix in which $P=\frac{1}{2}(I+AD^{-1})$.
Given the transition matrix $P$, we can calculate $x^{(t+1)}$ from
$x^{(t)}$ using the equation: 

\begin{equation}
x^{(t+1)}=Px^{(t)}.\label{eq:markov_chain_transition}
\end{equation}

A closed walk is a walk on a graph where the ending vertex is same
as the seed vertex. The period of a vertex is defined as the greatest
common divisor of the lengths of all closed walks that start from
this vertex. We say a graph is aperiodic if all of its vertices have
periods of 1. 

For an undirected, connected and aperiodic graph, there exists an
equilibrium state $\pi$, such that $\pi=P\pi$. This state is unique
and irrelevant to the starting point. By iterating Eq. \ref{eq:markov_chain_transition},
$x^{(t)}$ converges to $\pi$. More details about the Markov chain
process and the equilibrium state can be found from \cite{norris_markov_1998}.

\section{Limited Random Walk Procedure}

\subsection{Definitions}

We first define the transition matrix $P$. We assign the same probability
to the transition that the walking agent stays in the current vertex
and the transition that it moves to any neighboring vertex. We add
an identity matrix to the adjacency matrix and then normalize the
result to have each column sum to $1$. The transition matrix can
be written as 
\begin{equation}
P=\left(I+A\right)\left(I+D\right)^{-1}.\label{eq:eq_transition_matrix}
\end{equation}

Comparing to the transition matrix in Eq. \ref{eq:classifical_transition_matrix},
this is similar to adding self-loops to the graph. But we increase
the degree of each vertex by 1 instead of 2. This modification fixes
the periodicity problem that the graph may have. It greatly improves
the algorithm's stability and accuracy in graph clustering. 

At each walking step the probability vector $x^{(t)}$ is computed
using Eq. \ref{eq:markov_chain_transition}. Note that, in general,
elements in $x^{(t)}$ that are around the seed vertex are non-zeros
and the rest are zeros. So we do not need the full transition matrix
to calculate the probability vector for the next step.

Starting from the seed vertex, a normal random walk procedure will
eventually explore the whole graph. To reveal a local graph structure,
different techniques can be used to limit the scope of the walks.
In \cite{macropol_rrw:_2009} and \cite{andersen_using_2007}, the
random walk function is defined as 
\begin{equation}
x^{(t+1)}=\alpha x^{(0)}+(1-\alpha)Px^{(t)},\label{eq:page_rank_walk}
\end{equation}
 where $\alpha$ is called the teleport probability. The idea is that
there is a certain probability that the walking agent will teleport
back to the seed vertex and continue walking. 

Inspired by the Markov Clustering Algorithm (MCL) algorithm \cite{dongen_graph_2000},
we adapt the inflation and normalization operation after each step
of the transition. The inflation operation is an element-wise super-linear
function---a function that grows faster than a linear function. Here
we use the power function 

\begin{equation}
f\left(x\right)=\left[x_{1}^{r},x_{2}^{r},\ldots,x_{n}^{r}\right]^{T},\label{eq:eq_inflation}
\end{equation}
 where the exponent $r>1$. 

Since $x$ indicates the probability that the agent hits each vertex,
$x$ must be normalized to have a sum of 1 after the inflation operation.
The normalization function is defined as 

\begin{equation}
g(x)=\frac{x}{\left\Vert x\right\Vert _{1}},\label{eq:eq_normalization}
\end{equation}
where $\left\Vert x\right\Vert _{1}=\sum_{i=1}^{n}\left|x_{i}\right|$
is the $L_{1}$norm of the vector $x$. Since $x_{i}\ge0$ and $\sum_{i=1}^{n}x_{i}=1$,
Eq. \ref{eq:eq_normalization} can also be written in a vector form
as 
\begin{equation}
g(x)=\frac{x}{x^{T}\cdot\mathbf{1}},\label{eq:eq_normalisation_vector}
\end{equation}
 where $\mathbf{1}=\left[1,1,\ldots1\right]^{T}$. The inflation and
normalization operation enhance large values and depress small values
in the vector $x$. 

We call this procedure the Limited Random Walk (LRW) procedure. The
procedure limits the agent to walk around the neighborhood of the
seed vertex, especially if there is a clear graph cluster boundary. 

The MCL algorithm simulates flow within a graph. It uses the inflation
and normalization operation to enhance the flow within a cluster and
reduce the flow between clusters. The MCL procedure is a time-inhomogeneous
Markov Chain in which the transition matrix varies over time. The
MCL algorithm starts the random walk from all vertices simultaneously---there
are $n$ agents walking on the graph at the same time. The walking
can only continue after all agents have completed a walking step and
the result probability matrix has been inflated and normalized. Unlike
in the MCL algorithm, the LRW procedure is a time-homogeneous Markov
Chain. We initiate random walk from a single seed vertex, and do the
inflation on the probability values of this walking agent. This design
has many advantages. First, it avoids unnecessary walks since the
graph structure around the seed vertex may be exposed by a single
walk. Second, the procedure is suitable for the local clustering problems
because it does not require the whole graph data. Third, if multiple
walks are required, each walk procedure can be executed independently.
Thus the algorithm is fully parallelizable. 

The LRW procedure involves a nonlinear operation, thus it is difficult
to analyze its properties on a general graph model. Next we study
the equilibrium of the LRW procedure. Later, we will use these properties
to analyze the LRW procedure on general graphs.

\subsection{Equilibrium of the LRW procedure }

We first prove the existence of equilibrium of the LRW procedure.
Let $X$ be the set of values of the probability vector $x$. We have
\begin{eqnarray}
X & = & \left\{ (x_{1},x_{2},\cdots,x_{n})\vert0\le x_{1},x_{2},\cdots,x_{n}\le1\right.\nonumber \\
 &  & \left.\mathrm{and}\ x_{1}+x_{2}+\cdots x_{n}=1\right\} .\label{eq:lrw_function_domain}
\end{eqnarray}
 The LRW procedure defined by Eqs. \ref{eq:markov_chain_transition},
\ref{eq:eq_inflation} and \ref{eq:eq_normalization} is a function
that maps $X$ to itself. Let $\mathcal{L}:\ X\rightarrow X$, such
that 
\begin{equation}
\mathcal{L}(x)=g(f(Px)).\label{eq:lrw_function}
\end{equation}

\begin{thm}
\label{thm:lrw_stable_state}There exists a fixed-point $x^{*}$ such
that $\mathcal{L}\left(x^{*}\right)=x^{*}$. \end{thm}
\begin{IEEEproof}
We use the Brouwer fixed-point theorem to prove this statement. 

Given $0\le x_{1},x_{2},\cdots,x_{n}\le1$, the set X is clearly bounded
and closed. Thus $X$ is a compact set. 

Let $u,v\in X$ and $w=\lambda u+(1-\lambda)v$, where $\lambda\in\mathcal{R}$
and $0\le\lambda\le1$. So $w_{i}=\lambda u_{i}+(1-\lambda)v_{i}$
for $i=1,2,\cdots n$. Obviously $0\le w_{i}\le1$.

Further, 
\begin{eqnarray*}
\sum_{i=1}^{n}w_{i} & = & \sum_{i=1}^{n}\left(\lambda u_{i}+\left(1-\lambda\right)v_{i}\right)\\
 & = & \lambda\sum_{i=1}^{n}u_{i}+\left(1-\lambda\right)\sum_{i=1}^{n}v_{i}\\
 & = & 1
\end{eqnarray*}

Thus $w\in X$. This indicates that the set $X$ is convex. 

Since function $f(x)$ is continuous over the set $X$ and function
$g(x)$ is continuous over the codomain of function $f(x)$, function
$\mathcal{L}$ is continuous over the set $X$.

According to the Brouwer fixed-point theorem, there is a point $x^{*}$
such that $\mathcal{L}\left(x^{*}\right)=x^{*}$.
\end{IEEEproof}
Theorem \ref{thm:lrw_stable_state} shows the existence of fixed-point
of the LRW procedure, i.e., the LRW procedure will not escape from
a fixed-point whenever the point is reached. Since the LRW procedure
is a non-linear discrete dynamic system, it is difficult to analytically
investigate the system behavior. However, when $r=1$, the LRW procedure
is simply a Markov chain process, in which the fixed-point $x^{*}$
is the unique equilibrium state $\pi$ and the global attractor. In
another extreme case when $r\rightarrow\infty$, a fixed-point can
be an unstable equilibrium and the LRW procedure may have limit cycles
that oscillate around a star structure in the graph. In one state
of the oscillation, the probability value of the center of a star
structure is close to one. In practice, we chose $r$ from $(1,2]$,
where the LRW procedure is close to a linear system and oscillations
are extremely rare. In this case, the fixed-points of the LRW procedure
are stable equilibriums. 

Next we show how to use the LRW procedure to find clusters in a graph.

\subsection{Limited Random Walk on General Graphs\label{sub:Limited-Random-Walk}}

Without any prior knowledge of the cluster formation, we normally
start the LRW procedure from an initial state where $x_{s}=1$, $x_{i}=0$
for $i\ne s$ and $s$ is the seed vertex. During the LRW procedure,
there are two simultaneous processes---the spreading process and the
contracting process. When the two processes can balance each other,
the stationary state is reached. 

During the spreading process, the probability values spread as the
walking agent visits new vertices. The number of visited vertices
increases exponentially at first. The growth rate depends on the average
degree of the graph. The newly visited vertices will always receive
the smallest probability values. If the graph has an average degree
of $d$, it is not difficult to see that the expected probability
value of a newly visited vertex at step $t$ is $\left(\nicefrac{1}{d}\right)^{t}$.
As the walking continues, the probability values tend to be distributed
more evenly among all visited vertices. 

The other ongoing process during the LRW procedure is the contracting
process. During this process, the probability values of the visited
vertices contract to some vertices. Since the graph is usually heterogeneous,
some vertices (and groups of vertices) will receive higher probability
values as the procedure continues. The inflation operation further
enhances this contracting effect. The degrees of a vertex and its
surrounding vertices determine whether the probability values concentrate
to or diffuse from these vertices. Some vertices, normally the center
of a star structure, receive larger probability values than others.
We call these vertices feature vertices since they can be used to
represent the structure of a graph. 

Because the density of edge inside a cluster is higher than that of
linking the nodes inside and outside the cluster, the probability
that a walking agent visits nodes outside the cluster is small. Thus,
the LRW procedure will find feature vertices that the seed vertex
belongs to. We can use these vertices as features to cluster the vertices. 

The larger the inflation exponent $r$ is, the faster the algorithm
converges to the feature vertices. The LRW procedure tries to find
the feature vertices that are near the seed vertex. However, if $r$
is too large, the probability values concentrate to the nearest feature
vertex (or the seed vertex itself) before the graph is sufficiently
explored. If $r$ is too small, the probability values will concentrate
to the feature vertices that may belong to other clusters. The performance
of the LRW algorithm depends on choosing a proper inflation exponent
$r$. From this aspect, it is similar to the MCL algorithm. In practice,
$r$ is normally chosen between $1$ and $2$ and the value $2$ was
found to be suitable for most graphs.

\section{Limited Random Walk for Graph Clustering Problems}

\subsection{Global Graph Clustering Problems}

In this section we use the proposed LRW procedure in global graph
clustering problems. Our algorithm is divided into two phases---graph
exploring phase and cluster merging phase. To improve the performance
on big graph data, we also introduce a multi-stage strategy.

\subsubsection{\label{sub:Graph-Exploring-Phase}Graph Exploring Phase}

In the graph exploring phase, the LRW procedure is started from several
seed vertices. At each iteration, the agent moves one step as defined
in Eq. \ref{eq:markov_chain_transition}. Then the probability vector
$x$ is inflated by Eq. \ref{eq:eq_inflation} and normalized by Eq.
\ref{eq:eq_normalization}. The iteration stops when the probability
vector $x$ converges or the predefined maximum number of iterations
is reached. Let $x^{(*,i)}$ denote the final probability vector of
a random walk that was started from the seed vertex $v_{i}$. As described
in the previous section, the LRW procedure explores the vertices that
are close to the seed vertex. Thus, the vector $x^{(*,i)}$ has non-zero
elements only on these neighboring vertices. 

Algorithm \ref{alg:alg_lrw_exploring} illustrates the graph exploring
from a seed vertex set $Q$. Note that for small graph data, we can
set the seed vertex set $Q=V$ (i.e. the whole graph). In such case,
the LRW procedure is executed on every vertex of the graph and the
multi-stage strategy is not used.

\medskip{}

\begin{algorithm}[h]
\textbf{given} adjacency matrix $A$ of the graph $G(V,E)$, the seed
vertex set $Q$, the exponent $r$ of the inflation function \ref{eq:eq_inflation},
maximum number of iterations $T_{max}$, a small value $\epsilon$
to limit the number of visited vertices, and a small value $\xi$
for the termination condition

\textbf{initialize} the transition matrix $P$ by Eq. \ref{eq:eq_transition_matrix}

\textbf{for each }vertex $i$ in the vertex set $Q$
\begin{enumerate}
\item initialize $x^{(0,i)}$ such that $x_{k}^{(0,i)}=\begin{cases}
1 & k=i\\
0 & \mathrm{otherwise}
\end{cases}$ and $t=1$
\item repeat if $t<T_{max}$

\begin{enumerate}
\item $x^{(t,i)}=Px^{(t-1,i)}$
\item if $x_{k}^{(t,i)}<\epsilon$, set $x_{k}^{(t,i)}$ to 0
\item $x_{k}^{(t,i)}=f(x_{k}^{(t,i)})=\left(x_{k}^{(t,i)}\right)^{r}$
\item $x_{k}^{(t,i)}=\frac{x_{k}^{(t,i)}}{\left\Vert x^{(t,i)}\right\Vert _{1}}$
\item if $\left\Vert x^{(t,i)}-x^{(t-1,i)}\right\Vert _{2}<\xi$, exit the
loop
\end{enumerate}
\item output $x^{(t,i)}$ as the feature vector $x^{(*,i)}$ for vertex
$i$
\end{enumerate}
\protect\caption{Limited random walk graph exploring from a vertex set\label{alg:alg_lrw_exploring}}
\end{algorithm}

Note that the threshold $\epsilon$ limit the number of nonzero elements
in the probability vector $x$. It is easy to prove that the number
of nonzero elements in $x^{(t,i)}$ is less than $\nicefrac{1}{\epsilon}$.
A larger $\epsilon$ eliminates very small values in $x^{(t,i)}$
and prevent unnecessary computing efforts. However, $\epsilon$ does
not impose a limit on the largest cluster we can find. Further, the
choice of $\epsilon$ has little impact on the final clustering results
because either the LRW procedure finds the most dominant feature vertices
in a cluster or the small clusters are merged in the cluster merging
phase.

\subsubsection{\label{sub:Cluster-Merging-Phase}Cluster Merging Phase}

After the graph has been explored, we will find the clusters in the
cluster merging phase. We treat each $x^{(*,i)}$ as the feature vector
for the vertex $v_{i}$. Vertices belonging to the same cluster have
feature vectors that are close to each other. Any unsupervised clustering
algorithm, such as k-means or single linkage clustering method, can
be applied to find the desired number of ($k$) clusters. Because
of the computational complexity of these clustering algorithms, we
design a fast merging algorithm that can efficiently cluster vertices
according to their feature vectors. 

Each element $x_{j}^{(*,i)}$ in $x^{(*,i)}$ is the probability value
of the stationary state that the walking agent hits the vertex $v_{j}$
when the seed vertex is $v_{i}$. The feature vector $x^{(*,i)}$
is determined by the graph structure of the cluster that the initial
vertex $v_{i}$ belongs to. Thus, vertices in the same cluster should
have very similar feature vectors. We first find the vertex that has
the largest value in the vector $x^{(*,i)}$. Suppose $m=\arg\max_{j}\left(x_{j}^{(*,i)}\right)$,
we call $v_{m}$ the attractor vertex of vertex $v_{i}$. Grouping
vertices by their attractor vertex can be done in a fast way (complexity
of $O\mbox{(1)}$) using a dictionary data structure. After the grouping,
each vertex is assigned to a cluster that is identified by the attractor
vertex. However, it is possible that some vertices in one cluster
do not belong to the same attractor vertex. This may happen when the
cluster is large and the edge density in the cluster is low. We then
apply the following cluster merging algorithm to handle this overclustering
problem. 

The vertices that have large values, which determined by a threshold
relative to $x_{m}^{(*,i)}$, in $x^{(*,i)}$ are called significant
vertices for vertex $v_{i}$. If two vertices have large enough overlaps
of their significant vertices, they should be grouped into the same
cluster. From this observation, we first collect significant vertices
for the found clusters. Then we merge clusters if their significant
vertices overlap more than a half. Details of determining the significant
vertices and merging clusters are given in 

Note that the attractor vertex and the significant vertices are always
in the same cluster as the seed vertex. This is very useful when we
use the multi-stage graph strategy.

Algorithm \ref{alg:alg_lrw_merging} shows the details of the merging
phase of the LRW algorithm. Note that, for small graph data, we set
the seed vertex set Q=V and the initial clustering dictionary $\mathcal{D}$
to be empty. 

\medskip{}

\begin{algorithm}[h]
\textbf{given} feature vectors $x_{i}^{(*,i)}$ of the seed vertex
set $Q$, where $i\in Q$, threshold $\tau$ such that $0<\tau<1$
, and initial clustering dictionary $\mathcal{D}$

\textbf{for each} feature vector $x_{i}^{(*,i)}$ where $i\in Q$
\begin{enumerate}
\item find $m=\arg\max_{j}(x_{j}^{(*,i)})$ and add $i$ to an empty vertex
set $\mathcal{S}$
\item find all $j$ such that $x_{j}^{(*,i)}>\tau\cdot x_{m}^{(*,i)}$ and
add them to an empty vertex set $\mathcal{F}$
\item \textbf{if} $\mathcal{D}$ does not contain the key $m$

\quad{}add the pair $\left\langle \mathcal{S},\mathcal{F}\right\rangle $
as the value that associated with the key $m$ to dictionary $\mathcal{D}$

\textbf{else}

\quad{}find the value $\left\langle \mathcal{S}_{m},\mathcal{F}_{m}\right\rangle $
that is associated with the key $m$

\quad{}add $i$ to set $\mathcal{S}_{m}$ and merge $\mathcal{F}$
to $\mathcal{F}_{m}$ by $\mathcal{F}_{m}=\mathcal{F}_{m}\cup\mathcal{F}$

\quad{}update the value $\left\langle \mathcal{S}_{m},\mathcal{F}_{m}\right\rangle $
of the key $m$ in the dictionary $\mathcal{D}$.

\textbf{end}

\end{enumerate}
\textbf{for each }pair of keys $m_{1}$ and $m_{2}$ in the dictionary
$\mathcal{D}$ 
\begin{enumerate}
\item get the value pairs of $\left\langle \mathcal{S}_{m_{1}},\mathcal{F}_{m_{1}}\right\rangle $
and $\left\langle \mathcal{S}_{m_{2}},\mathcal{F}_{m_{2}}\right\rangle $
that are associated with the key $m_{1}$ and $m_{2}$
\item get the union of the significant vertex set of the two clusters by
$\mathcal{U}=\mathcal{F}_{m_{1}}\cap\mathcal{F}_{m_{2}}$
\item if $\left|\mathcal{U}\right|>\frac{1}{2}\min\left(\left|\mathcal{F}_{m_{1}}\right|,\left|\mathcal{F}_{m_{2}}\right|\right)$

\begin{enumerate}
\item merge $\mathcal{S}_{m_{2}}$ to $\mathcal{S}_{m_{1}}$ , $\mathcal{S}_{m_{1}}=\mbox{\ensuremath{\mathcal{S}}}_{m_{1}}\cup\mathcal{S}_{m_{2}}$
\item merge $\mathcal{F}_{m_{2}}$ to $\mathcal{F}_{m_{1}}$ , $\mathcal{F}_{m_{1}}=\mathcal{F}_{m_{1}}\cup\mathcal{F}_{m_{2}}$
\item update $\left\langle \mathcal{S}_{m_{1}},\mathcal{F}_{m_{1}}\right\rangle $
to the key $m_{1}$
\item delete the key $m_{2}$ and its associated value
\end{enumerate}
\end{enumerate}
\textbf{for each} key $m$ in $\mathcal{D}$, output $\mathcal{S}_{m}$as
the clustering result

\protect\caption{LRW cluster merging phase\label{alg:alg_lrw_merging}}
\end{algorithm}

\subsubsection{Multi-stage Strategy}

For small graph data, we can do LRW procedure on every vertex of the
graph. So the seed vertex set $Q=V$. The graph clustering is completed
after a graph exploring phase and a cluster merging phase. However,
when the graph data is large, it is time-consuming to perform the
LRW procedure from every vertex of the graph. A multi-stage strategy
can be used to greatly reduce the number of required walkings. First,
we start the LRW procedure from a randomly selected vertex set. After
the first round of the graph exploring, some clusters can be found
after the cluster merging phase. Next we generate a new seed vertex
set by randomly selecting vertices from those vertices that have not
been clustered. Then we do the graph exploration from the new seed
vertex set. We repeat this procedure until all vertices are clustered. 

Algorithm \ref{alg:alg_lrw_multistage-exploring-1} shows the global
graph clustering algorithm using the multi-stage strategy.

\begin{algorithm}[h]
\textbf{given} adjacency matrix $A$ of the graph $G(V,E)$, the exponent
$r$ for the inflation function \ref{eq:eq_inflation}, threshold
value $\eta$, maximum number of iterations $T_{max}$ and a small
enough value $\epsilon$

\textbf{initialize} the transition matrix $P$ by Eq. \ref{eq:eq_transition_matrix},
vertex set $B=V$ and clustering dictionary $\mathcal{D}=\emptyset$

\textbf{while }B is not empty
\begin{enumerate}
\item generate a vertex set $Q\subset B$ by randomly select vertices from
$B$
\item do the graph exploring from the vertex set $Q$ using algorithm \ref{alg:alg_lrw_exploring}
to get feature vectors $x^{(*,i)}$ for $i\in Q$
\item given $x^{(*,i)}$ and $\mathcal{D}$, do the graph cluster merging
using algorithm \ref{alg:alg_lrw_merging} to update the clustering
dictionary $\mathcal{D}$
\item \textbf{for each }key $m$ in $\mathcal{D}$, merge the feature vertex
set $\mathcal{F}_{m}$ to the cluster vertex set $S_{m}$, $\mathcal{S}_{m}=\mbox{\ensuremath{\mathcal{S}}}_{m}\cup\mathcal{F}_{m}$
\item get clustered vertex set $R=\bigcup_{m\in\mathcal{D}}\mathcal{S}_{m}$ 
\item let $B=B\backslash R$
\end{enumerate}
\textbf{for each} key $m$ in $\mathcal{D}$, output $\mathcal{S}_{m}$as
the clustering result

\protect\caption{Global graph clustering algorithm using multi-stage strategy\label{alg:alg_lrw_multistage-exploring-1}}
\end{algorithm}

\subsection{Local Graph Clustering Problems}

For the local graph clustering problems, the LRW procedure can efficiently
find the cluster from a given seed vertex. To achieve this, we first
perform graph exploring from the seed vertex in the same way as described
in Section \ref{sub:Graph-Exploring-Phase}. Let $x^{(*)}$ be the
probability vector after the graph exploration. If a probability value
in $x^{(*)}$ is large enough, the corresponding vertex is assigned
the local cluster without further computation. Similar to the global
graph clustering algorithm, we use a relative threshold $\eta$ that
is related to the maximum value in $x^{(*)}$. Vertices whose probability
values are greater than $\eta\cdot\max\left(x_{j}^{(*)}\right)$ are
called significant vertices. The significant vertices are assigned
to the local cluster directly. A small value of $\eta$ will reduce
the computational complexity, but may decrease the accuracy of the
algorithm. Suitable values of $\eta$ were experimentally found to
be between $0.3$ and $0.5.$ 

The vertices with low probability values can either be outside of
the cluster or inside the cluster but with relatively low significance.
Unlike \cite{spielman_local_2008,chung_local_2013,macko_local_2013},
which involve a sweep operation and a cluster fitness function, we
do another round of graph exploring from these insignificant vertices.
After the second graph exploring is completed, we apply the cluster
merging algorithm described in Section \ref{sub:Cluster-Merging-Phase}. 

Algorithm \ref{alg:alg_lrw_local} presents the LRW local clustering
algorithm. 

\begin{algorithm}[h]
\textbf{given} graph $G(V,E)$, seed vertex $v$ and a threshold value
$\eta,$ where $0<\eta<1$ 
\begin{enumerate}
\item do graph exploration starting from the seed vertex $v$ as described
in algorithm \ref{alg:alg_lrw_exploring} and get the feature vector
$x^{(*,v)}$
\item find vertices in $x^{(*,v)}$ for which $x_{i}^{(*,v)}>0$ and collect
the vertices in to set $S$
\item find $m=\arg\max_{j}(x_{j}^{(*,v)})$ 
\item split the set $S$ into two sets $S_{1}$ and $S_{2}$ such that $S_{1}=\left\{ j|x_{j}^{(*,v)}\ge\eta x_{m}^{(*,v)}\right\} $
and $S_{2}=\left\{ j|x_{j}^{(*,v)}<\eta x_{m}^{(*,v)}\right\} $
\item for each vertex in $S_{2}$ do graph exploration to get feature vectors
$x^{(*,i)}$, where $i\in S_{2}$
\item do clustering merging according to the feature vectors of the vertex
$v$ and the vertices in $S_{2}$ as described in algorithm \ref{alg:alg_lrw_merging}
and find the cluster $S_{3}$ that contains the seed vertex $v$
\item return $S_{1}\cup S_{3}$
\end{enumerate}
\protect\caption{LRW local graph clustering algorithm\label{alg:alg_lrw_local}}
\end{algorithm}

\subsection{Computational Complexity}

We first analyze the computational complexity of the LRW algorithm
for the global graph clustering problem. We assume the graph $G(V,E)$
has clusters. Let $\bar{n}_{c}$ be the average cluster size---the
number of vertices in the cluster, and $C$ is the number of clusters.
We have $\bar{n}_{c}\cdot C=n$. Note $C\ll n$. The most time-consuming
part of the algorithm is the graph exploring phase. For each vertex,
every iteration involves a multiplication of the transition matrix
$P$ and the probability vector $x$. The LRW procedure visits not
only the vertices in the cluster but also a certain amount of vertices
close to the cluster. Let $\gamma$ be the coefficient that indicates
how far the LRW procedure explores the graph before it converges.
Notice the maximum number of nonzero elements in a probability vector
is $\nicefrac{1}{\epsilon}$. Let $J$ denote the number of vertices
that the LRW procedure visits in each iteration, thus $J=\min\left(\gamma\bar{n}_{c},\nicefrac{1}{\epsilon}\right)$.
Thus the transition step at each iteration has complexity of $O\left(J\bar{n}_{c}\right)$.
The inflation and normalization steps, which operate on the probability
vector $x$, have the complexity of $O\left(J\right)$. Let $K$ be
the number of iterations for the LRW procedure to converge. So, the
computational complexity for a complete LRW procedure on each vertex
is $O(KJ\bar{n}_{c})$. For a global clustering problem when performing
the LRW procedure on every vertex, the graph exploration phase has
a complexity of $O\left(KJ\bar{n}_{c}n\right)$. In the worst case,
the algorithm has a complexity of $O\left(n^{3}\right)$. This is
an extremely rare case and it only happens when the graph is small;
does not have a cluster structure; and the edge density is high. This
worst case scenario is identical to the MCL algorithm \cite{dongen_graph_2000}.
Notice that the variables $J$ and $K$ have upper bounds and $\bar{n}_{c}$
is determine by the graph structure, the algorithm has a complexity
of $O(n)$ for big graph data. 

The computational complexity of the cluster merging phase involves
merging clusters that were found using the attractor vertices. This
merging requires ${C \choose 2}$ sets comparison operations, where
$C$ is the number of clusters found by the attractor vertices. The
complexity of this phase is roughly $O(C^{2})$. This does not impose
a significant impact to the overall complexity of the algorithm, since
$C\ll n$. The time spent in this phase is often negligible. Experiments
show that the clusters found using the attractor vertices are close
to the final results. For applications where speed is more important
than accuracy, the cluster merging phase can be left out. 

When the LRW algorithm is used in local graph clustering problem,
the first graph exploration (started from the seed vertex) has a complexity
of $O(KJ\bar{n}_{c})$. After the first graph exploration, there are
$LJ$ vertices to be further explored, where $L$ is related to the
threshold $\eta$ and $L<1$. The overall complexity of the LRW local
clustering algorithm is thus $O(LKJ^{2}\bar{n}_{c})$. 

The LRW algorithm is a typical example of embarrassingly parallel
paradigm. In the graph exploring phase, each random walk can be executed
independently. Therefore it can be entirely implemented in a parallel
computing environment such as a high-performance computing system.
The time spent for graph exploring phase decreases roughly linearly
with respect to the number of available computing resources. The two-phase
design also fits the MapReduce programming model and can easily be
adapted into any MapReduce framework \cite{dean_mapreduce:_2008}.

\section{Experiments}

The LRW algorithm uses the following parameters: inflation exponent
$r$, maximum number of iterations $T_{max}$, small value $\epsilon$,
merging threshold $\tau$ and local clustering threshold $\eta$.
In practice, except the inflation exponent $r$, the values of the
other parameters have little impact to the final results. The inflation
power $r$ should be chosen according to the density of the graph.
A sparse graph should use a smaller value of $r$, though $r=2$ is
suitable for most real world graphs. In our experiments, we chose
$r=2$ unless otherwise specified. The other parameters have been
set as: $T_{max}=100,$ $\epsilon=10^{-5}$ and $\tau=\eta=0.3$.
We will show the impact of some parameters in Section \ref{sub:The-Impact-of_parameters}.

\subsection{Simulated Data for Global Graph Clustering Problem\label{sub:Simulated-Data-global}}

We first show the performance of the LRW algorithm using simulated
graph data. The simulated graph is generated using the Erdos-Renyi
model \cite{newman_networks:_2010} with some modifications to generate
clusters. Using the ground truth of the cluster structure, we can
evaluate the performance of graph clustering algorithms. This kind
of simulated data are widely used in the literature \cite{newman_fast_2004,danon_effect_2006,lancichinetti_benchmark_2008,newman_finding_2004}. 

The graphs are generated by the model $G(n,p,c,q)$ where $c$ is
the number of clusters, $n$ is the number of vertices, $p$ is the
probability of the link between two vertices, and $q=d_{in}/d_{out}$
is the parameter that indicates the strength of the cluster structure,
where $d_{in}$ is the expected number of edges linking one vertex
to other vertices inside the same cluster, and $d_{out}$ is the expected
number of edges linking a given vertex to other vertices in other
clusters. Larger $q$ indicates stronger cluster structure. When $q=1$,
each vertex has equal probability that it links to vertices that are
inside and outside of the cluster---the graph has a very weak cluster
structure. Let $d$ be the expected of degree of a vertex. So, $d=d_{in}+d_{out}=p(n-1)$.
We use this model to generate graphs that consist of $c$ clusters
and each cluster has the same number of vertices. For each pair of
vertices, we link them with the probability $\frac{qpc(n-1)}{(q+1)(n-c)}$
if they belong to the same cluster, and the probability of $\frac{pc(n-1)}{n(q+1)(c-1)}$
if they belong to different clusters. 

We use the normalized mutual information (NMI) to evaluate the clustering
result against the ground truth \cite{ana_robust_2003,danon_comparing_2005}.
We first calculate the confusion matrix where each row is a cluster
found by the clustering algorithm and each column is a cluster in
the ground truth. The entries in the confusion matrix are the cardinality
of the intersect set of the row cluster and the column cluster. Let
$N_{ij}$ be the value of the $i$-th row and $j$-th column, $N_{i-}$
the sum of the $i$-th row, $N_{-j}$ the sum of the $j$-th column,
$N$ the total number of vertices, $C_{A}$ the number of clusters
that the clustering algorithm found (number of rows), and $C_{G}$
the number of clusters in the ground truth (number of columns). The
NMI is calculated as follows: 

\begin{equation}
NMI=\frac{-2\sum_{i=1}^{C_{A}}\sum_{j=1}^{C_{G}}N_{ij}\log\left(N_{ij}N/N_{i-}N_{-j}\right)}{\sum_{i=1}^{C_{A}}N_{i-}\log\left(N_{i-}/N\right)+\sum_{j=1}^{C_{G}}N_{-j}\log\left(N_{-j}/N\right)},\label{eq:eq_nmi}
\end{equation}
where $0\le NMI\le1$.

 If the clustering algorithm returns the exact same cluster structure
as the ground truth, $NMI=1$. Notice, NMI is not a symmetric evaluation
metric. If an algorithm assigns all vertices into one cluster ($C_{A}=1$),
then NMI value is $0$. On the other hand if an algorithm assigns
each vertex to its own cluster ($C_{A}=N$), then $NMI>0$. 

\begin{figure}[h]
\centering{}%
\begin{tabular}{cc}
\includegraphics[bb=0bp 0bp 571bp 545bp,scale=0.2]{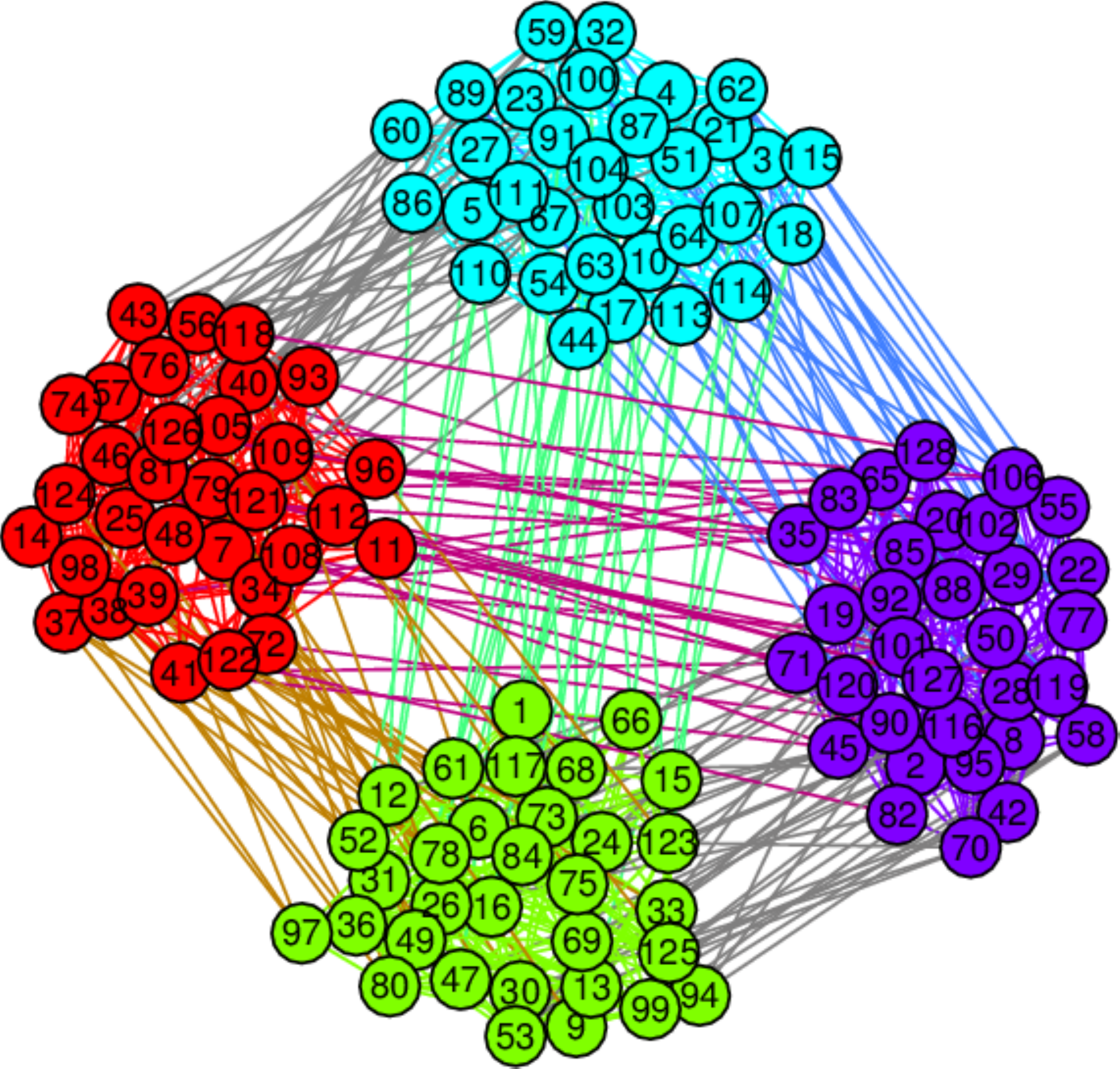} & \includegraphics[bb=0bp 0bp 562bp 534bp,scale=0.2]{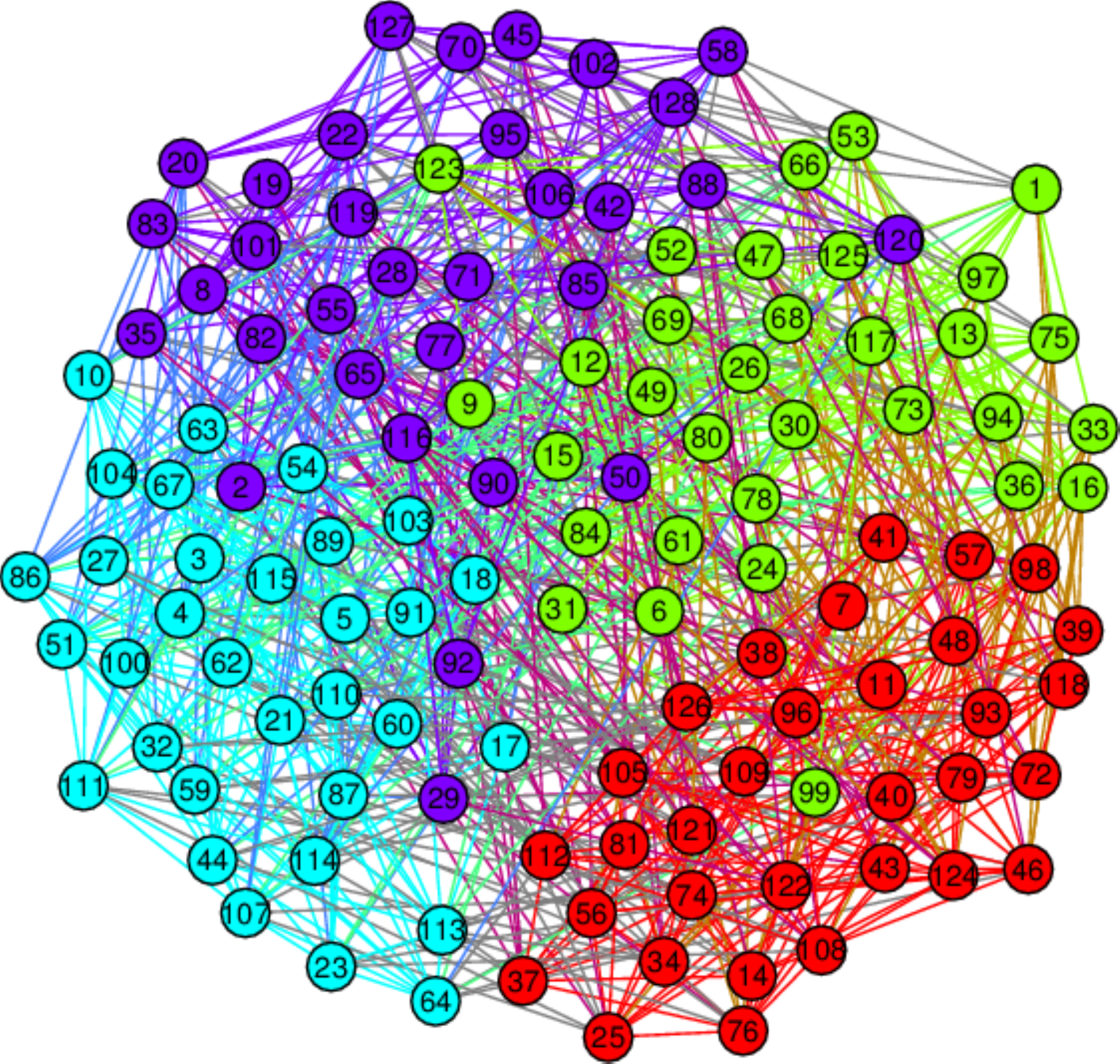}\tabularnewline
(a) & (b)\tabularnewline
\end{tabular} \protect\caption{\label{fig:fig_simdata_4}Simulated graphs. (a) $q=4$, (b) $q=1.22$}
\end{figure}

We generated graphs by choosing $n=128$ and $d=16$. The number of
the generated clusters is 4 and each cluster contains 32 vertices.
We varied the ratio $q$ and evaluated the performance of the LRW
algorithm against Girvan-Newman (GN) \cite{newman_finding_2004},
Louvain \cite{blondel_fast_2008}, Infomap \cite{rosvall_maps_2008}
and MCL \cite{dongen_graph_2000} algorithms. Two simulated graphs
are shown in Fig. \ref{fig:fig_simdata_4}, where the clusters are
colored differently and the graphs are visualized by force-directed
algorithms.

The comparative results are given in Table \ref{tbl:simulated_data},
where the number of clusters found by the algorithms are placed between
parentheses. 

\begin{table}[tbh]
\protect\caption{The NMI Values and the Numbers of Clusters of the Clustering Results
on Simulated Graph Data\label{tbl:simulated_data}}

\centering{}\renewcommand{\arraystretch}{1.3}%
\begin{tabular}{cccccc}
\hline 
q & GN & Louvain & Infomap & MCL & LRW\tabularnewline
\hline 
4.0 & 0.975 (4) & 1.0 (4) & 1 (4) & 1 (4) & 1.0 (4)\tabularnewline
3.0 & 1 (4) & 1.0 (4) & 1 (4) & 1 (4) & 1.0 (4)\tabularnewline
2.33 & 0.950 (4) & 1.0 (4) & 1 (4) & 0.860 (7) & 1.0 (4)\tabularnewline
1.86 & 0.900 (4) & 1.0 (4) & 1 (4) & 0.478 (95) & 1.0 (4)\tabularnewline
1.5 & 0.890 (4) & 1.0 (4) & 0 (1) & 0.453 (119) & 0.975 (4)\tabularnewline
1.22 & 0.593 (4) & 0.771 (5) & 0 (1) & 0.444 (128) & 0 (1)\tabularnewline
1 & 0.232 (4) & 0.304 (7) & 0 (1) & 0.444 (128) & 0 (1)\tabularnewline
\hline 
\end{tabular}
\end{table}

From the results, Louvain the best performing algorithm and LRW algorithm
comes as the second. It can be seen that the LRW algorithm can find
the correct structure if the graph has a strong clustering structure.
When the clustering structure diminishes as $q$ decreases, the LRW
algorithm returns the whole graph as one cluster. This behavior is
beneficial when we need to find the true clusters in a big graph.
The GN and Louvain algorithms are more like graph partition algorithms.
They optimize certain cluster fitness functions using the whole graph
data. They tend to partition the graph into clusters even though the
cluster structure is weak. 

We use real graph data to evaluate the performance of the LRW global
clustering algorithm on heterogeneous graphs. Details of the experiments
and the results are in Section \ref{sub:Real-World-Data}.

\subsection{Simulated Data for Local Graph Clustering Problems\label{sub:Simulated-Data-local}}

In this section, we compare the LRW algorithm with other local clustering
algorithms. The test graphs are generated using the protocol defined
in \cite{lancichinetti_benchmark_2008}. To simulate the data that
are close to real world graphs, the vertex degree and the cluster
size are chosen to follow the power law. Each test graph contains
2048 nodes. The vertex degree has the minimum value of 16 and the
maximum value of 128. The minimum and maximum cluster size are 16
and 256, respectively. Similar to the previous section, the inbound-outbound
ratio $q$ defines the strength of the cluster structure. 

The competing algorithms are criteria-based algorithms that optimize
a fitness function using either the greedy search or the simulated
annealing optimization method. Let $\mathcal{C}$ be the cluster that
contains the seed vertex. $\bar{\mathcal{C}}=V\backslash\mathcal{C}$
is the complement vertex set of $\mathcal{C}$. Let function $a(\cdot)$
be the total degree of a vertex set, that is 
\begin{equation}
a\left(S\right)=\sum_{i\in S,j\in V}A_{ij},
\end{equation}
where $A_{ij}$ are the entries of the adjacency matrix. The cut of
the cluster $\mathcal{C}$ is defined as 
\begin{equation}
c\left(\mathcal{C}\right)=\sum_{i\in\mathcal{C},j\in\bar{\mathcal{C}}}A_{ij}.
\end{equation}

The following are the definitions of the fitness functions. 

Cheeger constant (conductance): 
\begin{equation}
f(\mathcal{C})=\frac{c\left(\mathcal{C}\right)}{\min\left(a\left(\mathcal{C}\right),a\left(\bar{\mathcal{C}}\right)\right)}\label{eq:equation_conductance}
\end{equation}

Normalized cut: 
\begin{equation}
f(\mathcal{C})=\frac{c\left(\mathcal{C}\right)}{a\left(\mathcal{C}\right)}+\frac{c\left(\mathcal{C}\right)}{a\left(\bar{\mathcal{C}}\right)}
\end{equation}

Inverse relative density: 
\begin{equation}
f(\mathcal{C})=\frac{\left|E\right|-a\left(\mathcal{C}\right)+c\left(\mathcal{C}\right)}{a\left(\mathcal{C}\right)-c\left(\mathcal{C}\right)}
\end{equation}

Different local clustering algorithms are used to find the cluster
that contains the seed vertex. The Jaccard index is used to evaluate
the performance of each algorithm. Let $\mathcal{C}$ be the set of
vertices that an algorithm finds and $\mathcal{T}$ be the ground
truth cluster that contains the seed vertex. The Jaccard index is
defined as 

\begin{equation}
J=\frac{\left|\mathcal{C}\cap\mathcal{T}\right|}{\left|\mathcal{C}\cup\mathcal{T}\right|}
\end{equation}

We generated 10 test graphs for each inbound-outbound ratio $q$.
From each generated graph, we randomly picked 20 vertices as seeds.
For each algorithm and each inbound-outbound ratio $q$, we computed
the Jaccard index for each seed and took the average of the 200 Jaccard
indices. The results are shown in Table \ref{tbl:simulated_data_local},
where ``Che'' stands for the Cheeger constant (conductance) fitness
function; ``NCut'' stands for the normalized cut fitness function;
``IRD'' stands for the inverse relative density; the ending letter
``G'' stands for the greedy search method; and the ending letter
``S'' stands for the simulated annealing method. 

\begin{table}[tbh]
\protect\caption{Jaccard Index of Local Graph Clustering Results on the Simulated Graphs\label{tbl:simulated_data_local}}

\centering{}\renewcommand{\arraystretch}{1.3}%
\begin{tabular}{cccccccc}
\hline 
q & CheG & CheS & NCutG & NCutS & IRDG & IRDS & LRW\tabularnewline
\hline 
4.0 & 0.753 & 0.840 & 0.752 & 0.820 & 0.753 & 0.830 & \textbf{0.945}\tabularnewline
3.0 & 0.671 & 0.812 & 0.671 & 0.801 & 0.671 & 0.798 & \textbf{0.927}\tabularnewline
2.33 & 0.668 & 0.774 & 0.668 & 0.758 & 0.668 & 0.776 & \textbf{0.880}\tabularnewline
1.86 & 0.593 & 0.650 & 0.593 & 0.681 & 0.593 & 0.684 & \textbf{0.823}\tabularnewline
1.5 & 0.492 & 0.630 & 0.493 & 0.629 & 0.492 & 0.609 & \textbf{0.660}\tabularnewline
1.22 & 0.444 & 0.544 & 0.437 & \textbf{0.549} & 0.444 & 0.529 & 0.504\tabularnewline
1 & 0.298 & 0.410 & 0.296 & \textbf{0.452} & 0.298 & 0.424 & 0.295\tabularnewline
\hline 
\end{tabular}
\end{table}

From the results, the LRW algorithm greatly outperforms other methods
when the graph has a clear cluster structure.

\subsection{\label{sub:Real-World-Data}Real World Data}

In this section, we evaluate the performance of the LRW algorithm
on some real world graph data.

\subsubsection{Zachary's Karate Club}

\begin{figure}[h]
\centering{}\includegraphics[bb=0bp 0bp 568bp 430bp,scale=0.3]{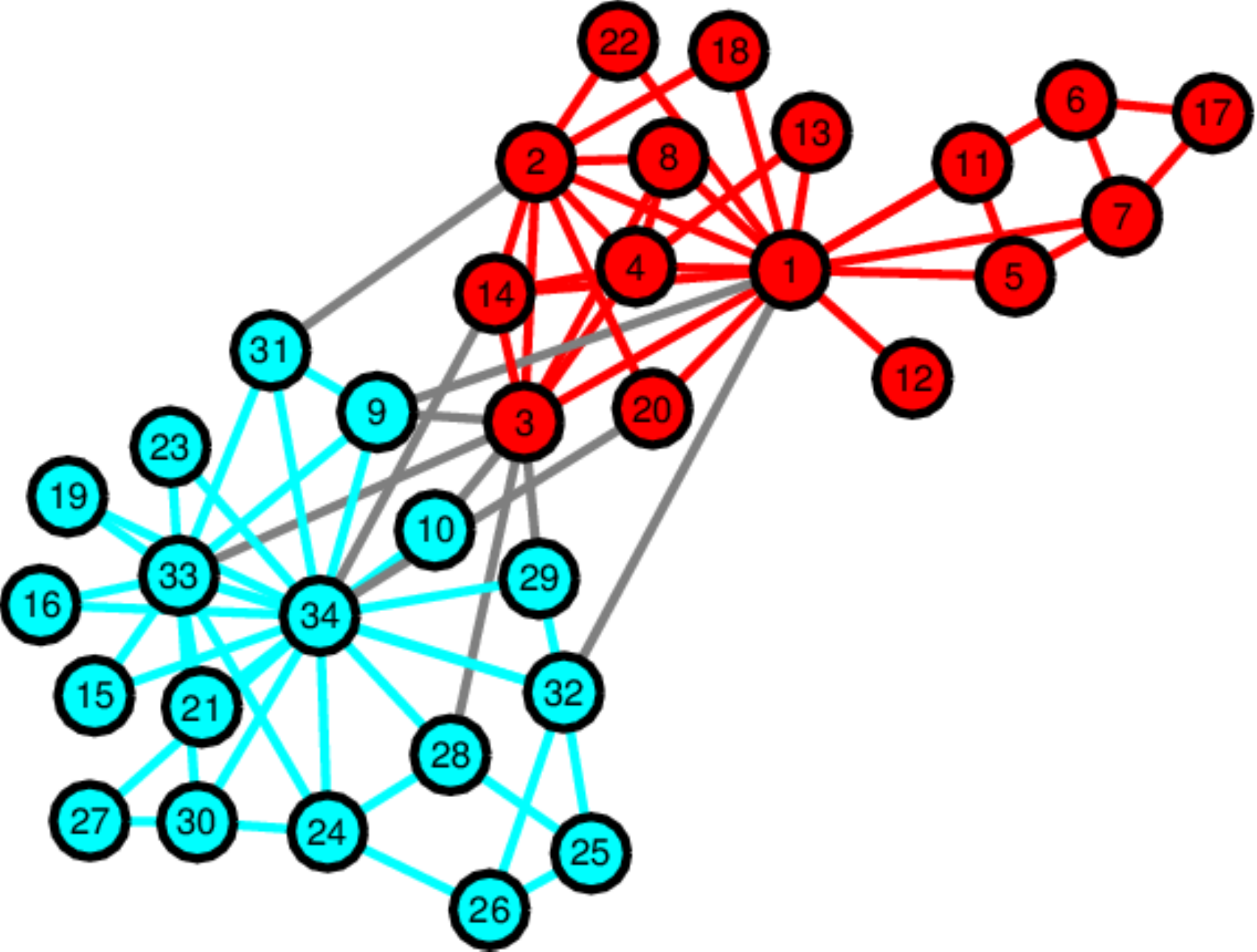}
\protect\caption{\label{fig:fig_karate_lrw}Clustering result of the karate club graph
data}
\end{figure}

We first do clustering analysis on the Zachary's karate club graph
data \cite{zachary_information_1977}. This graph is a social network
of friendship in a karate club in 1970. Each vertex represents a club
member and each edge represents the social interaction between the
two members. During the study, the club split into two smaller ones
due to the conflicts between the administrator and the coach. The
graph data have been regularly used to evaluate the performance of
the graph clustering algorithms \cite{newman_finding_2004,rosvall_maps_2008,sahai_hearing_2012}.
The graph contains 34 vertices and 78 edges. We applied the LRW algorithm
on this graph and the result shows two clusters that are naturally
formed. Fig. \ref{fig:fig_karate_lrw} shows the clustering result,
where clusters are illustrated using different colors. 

As the figure shows, the LRW algorithm finds the two clusters of the
Zachary's karate club. Actually the two clusters perfectly match the
ground truth---how the club was split in 1970. 

Clustering results of GN, Louvain, Infomap and MCL algorithms are
given in supplementary material.

\subsubsection{Ego-Facebook Graph Data\label{sub:Ego-Facebook-Graph-Data}}

The second data we used is the ego-Facebook graph data \cite{leskovec_learning_2012}.
The social network website Facebook allows users to organize their
friends into ``circles'' or ``friends lists'' (for example, friends
who share common interests). This data was collected from volunteer
Facebook users for researchers to develop automatic circle finding
algorithms. Ego-network is the network of an end user's friends. The
ego-Facebook graph is a combination of ego-networks from 10 volunteer
Facebook users. There are 4039 vertices and 88234 edges in the graph. 

We applied the LRW, GN \cite{newman_finding_2004}, Louvain \cite{blondel_fast_2008},
Infomap \cite{rosvall_maps_2008} and MCL \cite{dongen_graph_2000}
graph clustering algorithms to this data. To compare the results,
we generated the ground truth clustering by combining the vertices
in the ``circles'' of each volunteer user. So, the ground truth
contains 10 clusters. If a vertex appears in the circles of more than
one volunteer, we assign the vertex to all of these ground truth clusters.
We evaluated the number of clusters and the NMI values of the results
that each competing algorithm generates. 

We also calculated the mean conductance (MC) value of the clustering
results. The conductance value of a cluster is calculated using Eq.
\ref{eq:equation_conductance}. We then take the mean of all the conductance
values of the clusters that an algorithm finds. The smaller MC value
shows better clustering results. Note that MC tends to favor smaller
number of clusters in general. If the number of clusters are roughly
same, MC values give good evaluation of the clustering results. It
is also worth noting that MC value is capable of evaluating clustering
algorithms without the ground truth. We shall use this metric in later
experiments where the ground truth is not available.

The MC scores, NMI scores and the number of clusters found by each
algorithm are reported in Table \ref{tbl:ego-facebook_graph}. A bold
font indicates the best score among all competing algorithms.

\begin{table}[tbh]
\protect\caption{Global Graph Clustering Results on the Ego-Facebook Graph\label{tbl:ego-facebook_graph}}

\centering{}\renewcommand{\arraystretch}{1.3}%
\begin{tabular}{>{\centering}m{1.8cm}ccccc}
\hline 
 & GN & Louvain & Infomap & MCL & LRW\tabularnewline
\hline 
Mean Conductance & 0.156 & 0.133 & 0.397 & 0.0882 & \textbf{0.0770}\tabularnewline
NMI & 0.778 & 0.796 & 0.723 & 0.908 & \textbf{0.910}\tabularnewline
Number of Clusters & 16 & 19 & 76 & \textbf{10} & \textbf{10}\tabularnewline
\hline 
\end{tabular}
\end{table}

The results show that the random-walk-based algorithms---LRW and MCL---are
able to find the correct cluster structure of the data. Other criteria-based
algorithms are sensitive to trivial disparities of the graph structure
and are likely to overcluster the data.

\begin{figure}[h]
\centering{}\includegraphics[bb=0bp 0bp 561bp 418bp,scale=0.3]{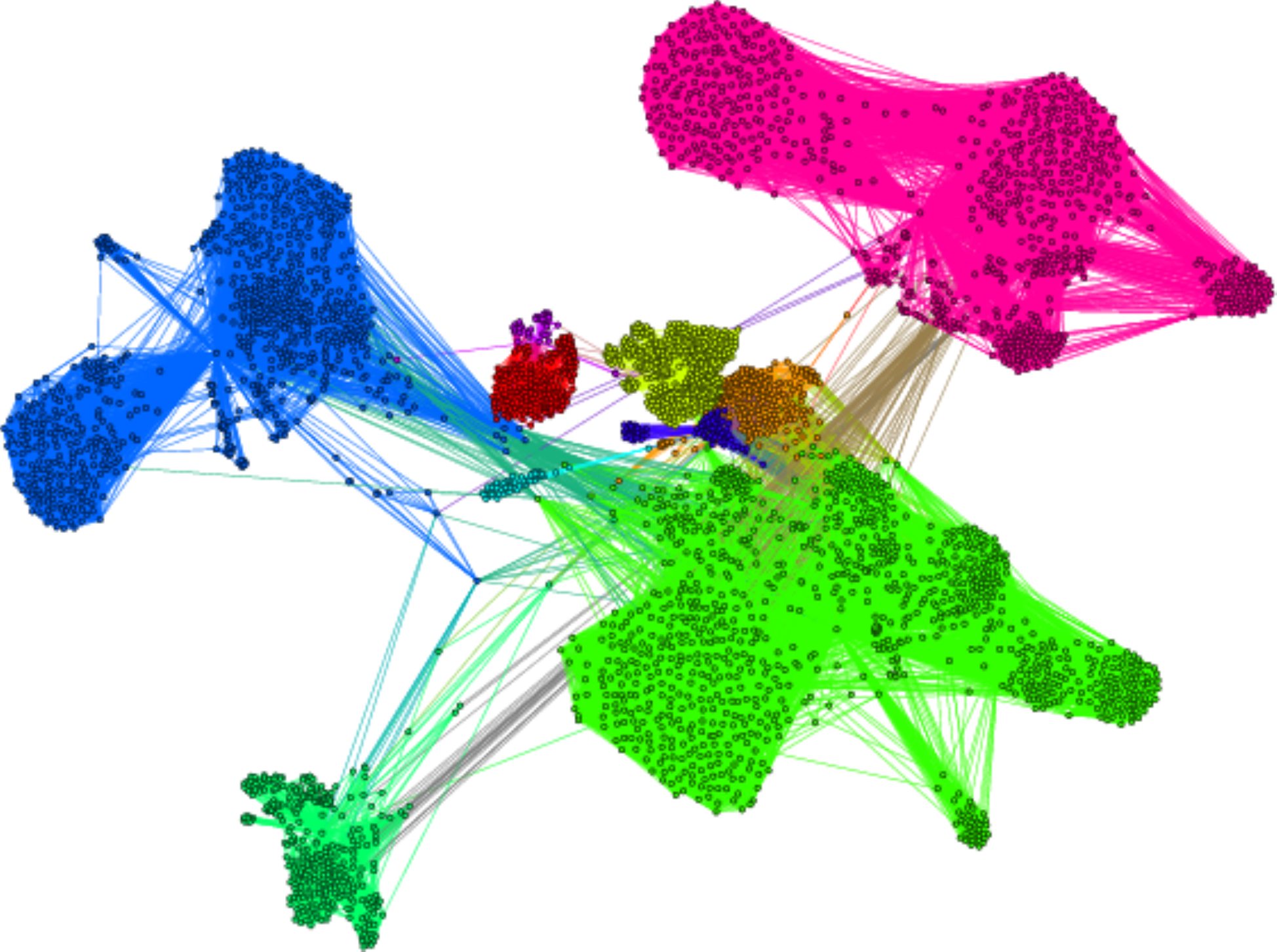}
\protect\caption{\label{fig:fig_facebook_lrw}Clustering result on the ego-Facebook
graph data}
\end{figure}

The clustering result of the LRW algorithms is shown in Fig. \ref{fig:fig_facebook_lrw}. 

Clustering results of other algorithms are shown in the supplementary
material.

\subsubsection{Heterogeneous Graph Data}

To evaluate the performance of the LRW algorithm on real heterogeneous
graph data, we selected 5 graph data from the collection of the KONECT
project \cite{kunegis_konect_2013}. The graph data are selected from
different categories and the size of the graph data varies from small
to medium. The properties and the references of the test graphs are
shown in Table \ref{tbl:real_network_graph}.

\begin{table}[tbh]
\protect\caption{Properties of the Testing Heterogeneous Graphs\label{tbl:real_network_graph}}

\centering{}\renewcommand{\arraystretch}{1.3}%
\begin{tabular}{>{\centering}m{1.6cm}cccc}
\hline 
 & Vertices & Edges & Category & Reference\tabularnewline
\hline 
dolphins & 62 & 156 & Animal & \cite{lusseau_bottlenose_2003}\tabularnewline
arenes-jazz & 198 & 2742 & Human Social & \cite{gleiser_community_2003}\tabularnewline
infectious & 410 & 2765 & Human Contact & \cite{isella_whats_2011}\tabularnewline
polblogs & 1490 & 19090 & Hyperlink & \cite{adamic_political_2005}\tabularnewline
reactome & 6229 & 146160 & Metabolic & \cite{croft_reactome_2013}\tabularnewline
\hline 
\end{tabular}
\end{table}

Since there is no ground truth available for these test data, we evaluated
each clustering algorithm by the mean conductance (MC) values. The
results are in Table \ref{tbl:real_network_conductance}. The best
MC score are shown in a bold font. The numbers of clusters found by
each algorithm are placed between parentheses. We also plot the clustered
graphs in which the vertices are located using a force-directed algorithm
and colored according to their associated clusters. These clustered
graphs are given in the supplementary material for subjective evaluation.

The reactome and the infectious graphs have low density. We chose
the inflation exponent $r=1.2$ to prevent overclustering the data.
For other graph data, the default value $r=2$ is used. 

\begin{table}[tbh]
\protect\caption{Global Graph Clustering Results on the Real Heterogeneous Graph Data\label{tbl:real_network_conductance}}

\centering{}\renewcommand{\arraystretch}{1.3}%
\begin{tabular}{>{\centering}m{1.3cm}>{\centering}m{1cm}>{\centering}m{1cm}>{\centering}m{1cm}>{\centering}m{1cm}>{\centering}m{1cm}}
\hline 
 & GN & Louvain & Infomap & MCL & LRW\tabularnewline
\hline 
dolphins & 0.425

(4) & 0.440

(5) & 0.487

(6) & 0.675

(12) & \textbf{0.347}

(4)\tabularnewline
arenes-jazz & 0.485

(4) & 0.455

(4) & 0.577

(7) & 0.529

(5) & \textbf{0.364}

(4)\tabularnewline
infectious  & \textbf{0.162}

(5) & 0.214

(6) & 0.465

(17) & 0.673

(40) & 0.175

(5)\tabularnewline
polblogs & 0.524

(12) & 0.501

(11) & 0.727

(36) & 0.777

(45) & \textbf{0.427}

(11)\tabularnewline
reactome  & 0.108

(110) & \textbf{0.099}

(114) & 0.315

(248) & 0.478

(352) & 0.221

(191)\tabularnewline
\hline 
\end{tabular}
\end{table}

Based on the MC scores and the visualized clustering results, the
LRW algorithm achieves a superior clustering performance in most of
the cases. Note that the reactome data has a weak cluster structure,
thus the LRW algorithm has difficulty to find a good partition for
it.

\subsection{The Sensitivity Analysis of the Parameters\label{sub:The-Impact-of_parameters}}

The proposed LRW algorithm depends on a number of parameters to perform
global and local graph clustering. In this section, we perform the
sensitivity analysis on the parameters. 

We use both simulated and real-world graphs in our experiments. Test
graph G1, G2 and G3 are similar to those used in Section \ref{sub:Simulated-Data-global}
except that we vary the density of each graph. The expected degree
$d$, which is a measure of the graph density, of graph G1, G2 and
G3 are 12, 16 and 20 respectively and $q$ is set to be 1.86 for all
test graphs. Test graph G4 is generated in the same way as described
in Section \ref{sub:Simulated-Data-local}. The ego-Facebook graph
data in Section \ref{sub:Ego-Facebook-Graph-Data} is used as an example
of real-world graphs. We performed global graph clustering on these
test graphs using the LRW algorithm with different parameters. NMI
scores are used to evaluate the performance of the algorithm. The
experiments using simulated graph data were repeated 10 times and
the average NMI scores and the average number of clusters are reported. 

As described in Section \ref{sub:Limited-Random-Walk}, the most important
parameter of the LRW algorithm is the inflation exponent $r$. We
first set $T_{max}=100$, $\epsilon=10^{-5}$, $\tau=0.3$ and vary
the inflation exponent $r$. Table \ref{tbl:parameters_different_r}
shows NMI scores and the number of clusters reported by the LRW algorithm
with different values of $r$. 

\begin{table}[tbh]
\protect\caption{NMI Scores and the Number of Clusters by the Different Inflation Exponent
Values\label{tbl:parameters_different_r}}

\centering{}\renewcommand{\arraystretch}{1.3}%
\begin{tabular}{>{\centering}m{1.3cm}>{\centering}m{1cm}>{\centering}m{1cm}>{\centering}m{1cm}>{\centering}m{1cm}>{\centering}m{1cm}}
\hline 
$r$ & G1 & G2 & G3 & G4 & Facebook\tabularnewline
\hline 
1.2 & 0

(1) & 0

(1) & 0

(1) & 0.155

(5) & 0.902

(10)\tabularnewline
1.4 & 0

(1) & 0

(1) & 0

(1) & 0.927

(22.7) & 0.906

(10)\tabularnewline
1.6 & 0

(1) & 0

(1) & 0

(1) & 1.0

(23) & 0.908

(11)\tabularnewline
1.8 & 1

(4) & 1

(4) & 1

(4) & 1.0

(20.8) & 0.910

(10)\tabularnewline
2 & 0.971

(4.6) & 1

(4) & 1

(4) & 1.0

(25) & 0.910

(10)\tabularnewline
2.4 & 0.868

(7.0) & 0.990

(4.2) & 1

(4) & 1.0

(21.8) & 0.910

(10)\tabularnewline
3 & 0.822

(6.3) & 0.962

(4.8) & 0.988

(4.2) & 1.0

(24.3) & 0.910

(10)\tabularnewline
\hline 
\end{tabular}
\end{table}

The test results show that the performance the LRW algorithm is sensitive
to the density of the test graphs. A large inflation exponent $r$
may overcluster the data as the results on graph G1 and G2 shows.
It can be easily noticed that the LRW algorithm is not sensitive to
the choice of $r$ for test graph G4 and the ego-Facebook graph. These
graphs, and almost all real-world graphs, are more heterogeneous than
the simulated graphs G1, G2 and G3. The LRW algorithm performs better
on this type of graph since the attractor vertices and significant
vertices are more stable on these graphs. 

The parameter $T_{max}$ sets a limit on the number of iterations
for the LRW procedure to converge. According to our experiments, value
100 is large enough to ensure the convergence of almost all cases.
For example, only 3 out of 88,234 LRW procedures do not converge within
100 iterations on the ego-Facebook graph. A few exceptional cases
has no impact on the final clustering results. The parameter $\epsilon$
is used to remove small values in the probability vector thus decrease
the computational complexity. It has no impact on the final clustering
result as long as the value is small enough, for example $\epsilon<10^{-4}$. 

We also conducted the experiments by varying the threshold value $\tau$
from $0.1$ to 0.5. The NMI scores and the number of clusters found
by the LRW algorithm with different $\tau$ values are almost identical
to those in Table \ref{tbl:parameters_different_r}. This indicates
that the choice of $\tau$ has very little impact on the clustering
performance. 

According to these results, one only needs to choose a proper inflation
exponent $r$ to use the LRW algorithms. Other parameters can be chosen
freely from a wide range of reasonable values. $r=2$ is suitable
for most of graphs and is preferable because of the computational
advantage.

\subsection{Big graph data}

In this section, we apply the LRW algorithm on real-world big graph
data and show the computational advantage of its parallel implementation.
The test graphs were received from the SNAP graph data collection
\cite{leskovec_stanford_????,yang_defining_2012}. These graphs are
from major social network services and E-commerce companies. We use
the high quality communities that either created by users or the system
as ground truth clusters. The details of the high quality communities
are described in \cite{yang_defining_2012}. The Rand Index is used
to evaluate the results of the proposed clustering algorithm. To generate
positive samples, we randomly picked 1000 pairs of vertices, where
the vertices in each pair come from the same cluster in the ground
truth. Negative samples consist of 1000 pairs of vertices, where the
vertices from each pair come from different clusters in the ground
truth. The Rand Index is defined as 
\begin{equation}
\mathrm{RI=}\frac{TP+TN}{N},
\end{equation}
 where $TP$ is the number of true positive samples, $TN$ is the
number of true negative samples, and $N$ is the total number of samples. 

Table \ref{tbl:big_graph_data} shows the size of the test graphs,
the time spent on the graph exploration phase, the number of CPU cores
and the amount of memory used for graph exploration, the time spent
on cluster merging phase, the number of clusters that the LRW algorithm
finds and the Rand Index of the clustering results. In this experiment,
multiple CPU cores were used for graph exploration and one CPU core
was used for clustering merging. Note that the competing algorithms
used in previous sections cannot complete this task due to the large
size of the data. 

\begin{table*}[tbh]
\protect\caption{Clustering Performance of Real-world Big Graph Data \label{tbl:big_graph_data}}

\centering{}\renewcommand{\arraystretch}{1.3}%
\begin{tabular}{>{\centering}m{5cm}>{\centering}m{2cm}>{\centering}m{2cm}>{\centering}m{2.5cm}>{\centering}m{2cm}}
\hline 
 & com-Amazon & com-Youtube & com-LiveJournal & com-Orkut\tabularnewline
\hline 
nodes & 334,863 & 1,134,890 & 3,997,962 & 3,072,441\tabularnewline
edges & 925,872 & 2,987,624 & 34,681,189 & 117,185,083\tabularnewline
CPU cores for graph exploration & 12 & 96 & 96 & 96\tabularnewline
memory per CPU core & 4G & 4G & 8G & 8G\tabularnewline
graph exploration (in hours) & 0.83 & 3.08 & 17.4 & 24.0\tabularnewline
cluster merging (in hours) & 0.20 & 1.34 & 9.44 & 2.53\tabularnewline
clusters & 37,473 & 170,569 & 381,246 & 165,624\tabularnewline
Rand Index & 0.908 & 0.755 & 0.951 & 0.751\tabularnewline
\hline 
\end{tabular}
\end{table*}

Table \ref{tbl:big_graph_data} shows that the LRW algorithm is able
to find clusters from large graph data with a reasonable computing
time and memory resource. The Rand Index values indicate that the
clusters returned by the LRW algorithm match well the ground truth.
The time spent on the graph exploration phase is inversely proportional
to the number of CPU cores. Computational time can be further reduced
if more computing resources are available. The proposed algorithm
can efficiently handle graphs with millions of nodes and hundreds
of millions of edges. For even larger graphs that exceed the memory
limit for each computing process, a mechanism that retrieves part
of the graph from a central storage can be used. Since the LRW procedure
is capable of exploring a limited number of vertices that are near
a seed vertex, the algorithm can cluster much larger graphs if such
a mechanism is implemented.

\section{Conclusions}

In this paper, we proposed a novel random-walk-based graph clustering
algorithm, the so-called LRW. We studied the behavior of the LRW procedure
and developed the LRW algorithms for both global and local graph clustering
problems. The proposed algorithm is fundamentally different from previous
random-walk-based algorithms. We use the LRW procedure to find attracting
vertices and use them as features to cluster vertices in a graph.
The performance of the LRW algorithm was evaluated using simulated
graphs and real-world big graph data. According to the results, the
proposed algorithm is superior to other well-known methods. 

The LRW algorithm can be efficiently used in both global and local
graph clustering problems. It finds clusters from a big graph data
by only locally exploring the graph. This is important for extreme
large data that may not even fit in a single computer memory. The
algorithm contains two phases---the graph exploring phase and the
cluster merging phase. The graph exploring phase is the most critical
part and also the most time-consuming part of the algorithm. This
phase can be implemented in embarrassingly parallel paradigm. The
algorithm can easily be adapted to any MapReduce framework. 

From our experiments, we also noticed the limitation of the LRW algorithm.
First, when used as a global clustering algorithm, the computational
complexity can be high, especially when the graph cluster structure
is weak. This is due to the fact that the graph may be analyzed multiple
times during the graph exploration phase, if we perform the LRW procedure
from every vertex of the graph. However, using the multi-stage strategy
can dramatically reduce the computation time. Second, if the cluster
structure is obscured, the LRW algorithm may return the whole graph
as one cluster---though this behavior is desired in many cases. 

The experiments show that the performance of the proposed LRW graph
clustering algorithm is not sensitive to any parameter except the
inflation exponent r, especially when the graph is not heterogeneous.
For future research, we will further improve the LRW algorithm so
that it can optimally select the inflation function that best suits
the problem at hand.

\appendices{}

\bibliographystyle{IEEEtran}
\bibliography{clean}

\begin{thebibliography}{10}
\providecommand{\url}[1]{#1}
\csname url@samestyle\endcsname
\providecommand{\newblock}{\relax}
\providecommand{\bibinfo}[2]{#2}
\providecommand{\BIBentrySTDinterwordspacing}{\spaceskip=0pt\relax}
\providecommand{\BIBentryALTinterwordstretchfactor}{4}
\providecommand{\BIBentryALTinterwordspacing}{\spaceskip=\fontdimen2\font plus
\BIBentryALTinterwordstretchfactor\fontdimen3\font minus
  \fontdimen4\font\relax}
\providecommand{\BIBforeignlanguage}[2]{{%
\expandafter\ifx\csname l@#1\endcsname\relax
\typeout{** WARNING: IEEEtran.bst: No hyphenation pattern has been}%
\typeout{** loaded for the language `#1'. Using the pattern for}%
\typeout{** the default language instead.}%
\else
\language=\csname l@#1\endcsname
\fi
#2}}
\providecommand{\BIBdecl}{\relax}
\BIBdecl

\bibitem{schaeffer_graph_2007}
S.~E. Schaeffer, ``Graph clustering,'' \emph{Computer Science Review}, vol.~1,
  no.~1, pp. 27--64, 2007.

\bibitem{newman_fast_2004}
M.~E. Newman, ``Fast algorithm for detecting community structure in networks,''
  \emph{Physical review E}, vol.~69, no.~6, p. 066133, 2004.

\bibitem{blondel_fast_2008}
V.~D. Blondel, J.-L. Guillaume, R.~Lambiotte, and E.~Lefebvre, ``Fast unfolding
  of communities in large networks,'' \emph{Journal of Statistical Mechanics:
  Theory and Experiment}, vol. 2008, no.~10, p. P10008, 2008.

\bibitem{clauset_finding_2004}
A.~Clauset, M.~E. Newman, and C.~Moore, ``Finding community structure in very
  large networks,'' \emph{Physical review E}, vol.~70, no.~6, p. 066111, 2004.

\bibitem{waltman_smart_2013}
L.~Waltman and N.~J.~v. Eck, ``\BIBforeignlanguage{en}{A smart local moving
  algorithm for large-scale modularity-based community detection},''
  \emph{\BIBforeignlanguage{en}{The European Physical Journal B}}, vol.~86,
  no.~11, pp. 1--14, Nov. 2013.

\bibitem{spielman_local_2008}
D.~A. Spielman and S.-H. Teng, ``A local clustering algorithm for massive
  graphs and its application to nearly-linear time graph partitioning,''
  \emph{arXiv preprint arXiv:0809.3232}, 2008.

\bibitem{qiu_graph_2006}
H.~Qiu and E.~R. Hancock, ``Graph matching and clustering using spectral
  partitions,'' \emph{Pattern Recognition}, vol.~39, no.~1, pp. 22--34, 2006.

\bibitem{spielman_spectral_2011}
D.~Spielman and S.~Teng, ``Spectral {Sparsification} of {Graphs},'' \emph{SIAM
  Journal on Computing}, vol.~40, no.~4, pp. 981--1025, Jan. 2011.

\bibitem{dongen_graph_2000}
S.~Dongen, ``Graph clustering by flow simulation,'' Ph.D. dissertation,
  Universiteit Utrecht, Utrecht, The Netherlands, May 2000.

\bibitem{macropol_rrw:_2009}
K.~Macropol, T.~Can, and A.~K. Singh, ``{RRW}: repeated random walks on
  genome-scale protein networks for local cluster discovery,'' \emph{BMC
  bioinformatics}, vol.~10, no.~1, p. 283, 2009.

\bibitem{chung_local_2013}
F.~Chung and M.~Kempton, ``A {Local} {Clustering} {Algorithm} for {Connection}
  {Graphs},'' in \emph{Algorithms and {Models} for the {Web} {Graph}}.\hskip
  1em plus 0.5em minus 0.4em\relax Springer, 2013, pp. 26--43.

\bibitem{macko_local_2013}
P.~Macko, D.~Margo, and M.~Seltzer, ``Local clustering in provenance graphs,''
  in \emph{Proceedings of the 22nd {ACM} international conference on
  {Conference} on information \& knowledge management}.\hskip 1em plus 0.5em
  minus 0.4em\relax ACM, 2013, pp. 835--840.

\bibitem{andersen_using_2007}
R.~Andersen, F.~Chung, and K.~Lang, ``Using pagerank to locally partition a
  graph,'' \emph{Internet Mathematics}, vol.~4, no.~1, pp. 35--64, 2007.

\bibitem{buhler_constrained_2013}
T.~B{\"u}hler, S.~S. Rangapuram, S.~Setzer, and M.~Hein, ``Constrained
  fractional set programs and their application in local clustering and
  community detection,'' \emph{arXiv:1306.3409 [cs, math, stat]}, Jun. 2013.

\bibitem{zhu_local_2013}
Z.~A. Zhu, S.~Lattanzi, and V.~Mirrokni, ``A local algorithm for finding
  well-connected clusters,'' in \emph{Proceedings of the 30th {International}
  {Conference} on {Machine} {Learning} ({ICML}-13)}, 2013, pp. 396--404.

\bibitem{norris_markov_1998}
J.~Norris, ``Markov {Chains} {\textbar} {Applied} probability and stochastic
  networks,'' 1998.

\bibitem{dean_mapreduce:_2008}
J.~Dean and S.~Ghemawat, ``{MapReduce}: {Simplified} {Data} {Processing} on
  {Large} {Clusters},'' \emph{Commun. ACM}, vol.~51, no.~1, pp. 107--113, Jan.
  2008.

\bibitem{newman_networks:_2010}
M.~Newman, \emph{\BIBforeignlanguage{English}{Networks: {An} {Introduction}}},
  1st~ed.\hskip 1em plus 0.5em minus 0.4em\relax Oxford ; New York: Oxford
  University Press, May 2010.

\bibitem{danon_effect_2006}
L.~Danon, A.~D{\'i}az-Guilera, and A.~Arenas, ``\BIBforeignlanguage{en}{The
  effect of size heterogeneity on community identification in complex
  networks},'' \emph{\BIBforeignlanguage{en}{Journal of Statistical Mechanics:
  Theory and Experiment}}, vol. 2006, no.~11, p. P11010, Nov. 2006.

\bibitem{lancichinetti_benchmark_2008}
A.~Lancichinetti, S.~Fortunato, and F.~Radicchi, ``Benchmark graphs for testing
  community detection algorithms,'' \emph{Physical Review E}, vol.~78, no.~4,
  Oct. 2008.

\bibitem{newman_finding_2004}
M.~E. Newman and M.~Girvan, ``Finding and evaluating community structure in
  networks,'' \emph{Physical review E}, vol.~69, no.~2, p. 026113, 2004.

\bibitem{ana_robust_2003}
L.~Ana and A.~Jain, ``Robust data clustering,'' in \emph{2003 {IEEE} {Computer}
  {Society} {Conference} on {Computer} {Vision} and {Pattern} {Recognition},
  2003. {Proceedings}}, vol.~2, Jun. 2003, pp. II--128--II--133 vol.2.

\bibitem{danon_comparing_2005}
L.~Danon, A.~Diaz-Guilera, J.~Duch, and A.~Arenas, ``Comparing community
  structure identification,'' \emph{Journal of Statistical Mechanics: Theory
  and Experiment}, vol. 2005, no.~09, p. P09008, 2005.

\bibitem{rosvall_maps_2008}
M.~Rosvall and C.~T. Bergstrom, ``Maps of random walks on complex networks
  reveal community structure,'' \emph{Proceedings of the National Academy of
  Sciences}, vol. 105, no.~4, pp. 1118--1123, Jan. 2008.

\bibitem{zachary_information_1977}
W.~W. Zachary, ``An information flow model for conflict and fission in small
  groups,'' \emph{Journal of anthropological research}, pp. 452--473, 1977.

\bibitem{sahai_hearing_2012}
T.~Sahai, A.~Speranzon, and A.~Banaszuk, ``Hearing the clusters of a graph: {A}
  distributed algorithm,'' \emph{Automatica}, vol.~48, no.~1, pp. 15--24, Jan.
  2012.

\bibitem{leskovec_learning_2012}
J.~Leskovec and J.~J. Mcauley, ``Learning to {Discover} {Social} {Circles} in
  {Ego} {Networks},'' in \emph{Advances in {Neural} {Information} {Processing}
  {Systems} 25}, F.~Pereira, C.~J.~C. Burges, L.~Bottou, and K.~Q. Weinberger,
  Eds.\hskip 1em plus 0.5em minus 0.4em\relax Curran Associates, Inc., 2012,
  pp. 539--547.

\bibitem{kunegis_konect_2013}
J.~Kunegis, ``{KONECT} {\textendash} {The} {Koblenz} {Network} {Collection},''
  in \emph{Proc. {Int}. {Conf}. on {World} {Wide} {Web} {Companion}}, 2013, pp.
  1343--1350.

\bibitem{lusseau_bottlenose_2003}
D.~Lusseau, K.~Schneider, O.~J. Boisseau, P.~Haase, E.~Slooten, and S.~M.
  Dawson, ``\BIBforeignlanguage{en}{The bottlenose dolphin community of
  {Doubtful} {Sound} features a large proportion of long-lasting
  associations},'' \emph{\BIBforeignlanguage{en}{Behavioral Ecology and
  Sociobiology}}, vol.~54, no.~4, pp. 396--405, Jun. 2003.

\bibitem{gleiser_community_2003}
P.~Gleiser and L.~Danon, ``Community {Structure} in {Jazz},'' \emph{Advances in
  Complex Systems}, vol.~06, no.~04, pp. 565--573, Dec. 2003.

\bibitem{isella_whats_2011}
L.~Isella, J.~Stehl{\'e}, A.~Barrat, C.~Cattuto, J.-F. Pinton, and W.~Van~den
  Broeck, ``What's in a {Crowd}? {Analysis} of {Face}-to-{Face} {Behavioral}
  {Networks},'' \emph{Journal of Theoretical Biology}, vol. 271, no.~1, pp.
  166--180, 2011.

\bibitem{adamic_political_2005}
L.~A. Adamic and N.~Glance, ``The {Political} {Blogosphere} and the 2004
  {U}.{S}. {Election}: {Divided} {They} {Blog},'' in \emph{Proceedings of the
  3rd {International} {Workshop} on {Link} {Discovery}}, ser. {LinkKDD}
  '05.\hskip 1em plus 0.5em minus 0.4em\relax New York, NY, USA: ACM, 2005, pp.
  36--43.

\bibitem{croft_reactome_2013}
D.~Croft, A.~F. Mundo, R.~Haw, M.~Milacic, J.~Weiser, G.~Wu, M.~Caudy,
  P.~Garapati, M.~Gillespie, M.~R. Kamdar, B.~Jassal, S.~Jupe, L.~Matthews,
  B.~May, S.~Palatnik, K.~Rothfels, V.~Shamovsky, H.~Song, M.~Williams,
  E.~Birney, H.~Hermjakob, L.~Stein, and P.~D'Eustachio,
  ``\BIBforeignlanguage{en}{The {Reactome} pathway knowledgebase},''
  \emph{\BIBforeignlanguage{en}{Nucleic Acids Research}}, p. gkt1102, Nov.
  2013.

\bibitem{leskovec_stanford_????}
J.~Leskovec, ``Stanford {Large} {Network} {Dataset} {Collection}.''

\bibitem{yang_defining_2012}
J.~Yang and J.~Leskovec, ``Defining and {Evaluating} {Network} {Communities}
  based on {Ground}-truth,'' \emph{arXiv:1205.6233 [physics]}, May 2012.

\end{thebibliography}

\begin{IEEEbiography}[{\fbox{\begin{minipage}[t][1.25in]{1in}%
\includegraphics[clip,width=1in,height=1.25in]{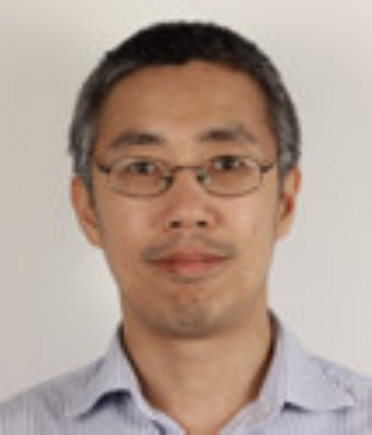}%
\end{minipage}}}]{Honglei Zhang}
is a PhD student and a researcher in the Department of Signal Processing
at Tampere University of Technology. He received his Bachelor and
Master degree in Electrical Engineering from Harbin Institute of Technology
in 1994 and 1996 in China, respectively. He worked as a software engineer
in Founder Co. in China from 1996 to 1999. He had been working as
a software engineer and a software system architect in Nokia Oy Finland
for 14 years. He has 6 scientific publications and 2 patents. His
current research interests include computer vision, pattern recognition,
data mining and graph algorithms. More about of his research can be
found from \url{http://www.cs.tut.fi/~zhangh}.
\end{IEEEbiography}

\begin{IEEEbiography}[{\fbox{\begin{minipage}[t][1.25in]{1in}%
\includegraphics[clip,width=1in,height=1.25in]{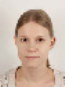}%
\end{minipage}}}]{Jenni~Raitoharju}
received her MS degree in information technology in 2009 from Tampere
University of Technology and she\textquoteright s now pursuing a PhD
degree. She is currently working as a researcher at the Department
of Signal Processing, Tampere University of Technology. Her research
interests include content-based image analysis, indexing and retrieval,
machine learning, neural networks, stochastic optimization and divide
and conquer algorithms.
\end{IEEEbiography}

\vfill{}

\newpage{}
\begin{IEEEbiography}[{\fbox{\begin{minipage}[t][1.25in]{1in}%
\includegraphics[clip,width=1in,height=1.25in]{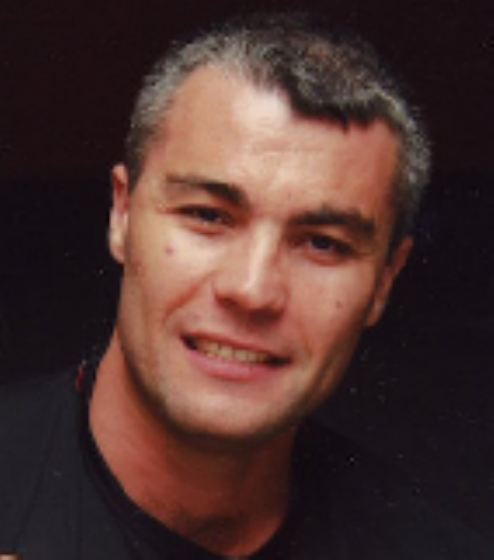}%
\end{minipage}}}]{Serkan Kiranyaz}
 was born in Turkey, 1972. He received his BS degree in Electrical
and Electronics Department at Bilkent University, Ankara, Turkey,
in 1994 and MS degree in Signal and Video Processing from the same
University, in 1996. He worked as a Senior Researcher in Nokia Research
Center and later in Nokia Mobile Phones, Tampere, Finland. He received
his PhD degree in 2005 and his Docency at 2007 from Tampere University
of Technology, respectively. He is currently a Professor in the Department
of Electrical Engineering at Qatar University. Prof. Kiranyaz published
2 books, more than 30 journal papers in several IEEE Transactions
and some other high impact journals and 70+ papers in international
conferences. His recent publication has been nominated for the Best
Paper Award in IEEE ICIP\textquoteright 13 conference. Another publication
won the IBM Best Paper Award in ICPR\textquoteright 14.
\end{IEEEbiography}

\begin{IEEEbiography}[{\fbox{\begin{minipage}[t][1.25in]{1in}%
\includegraphics[clip,width=1in,height=1.25in]{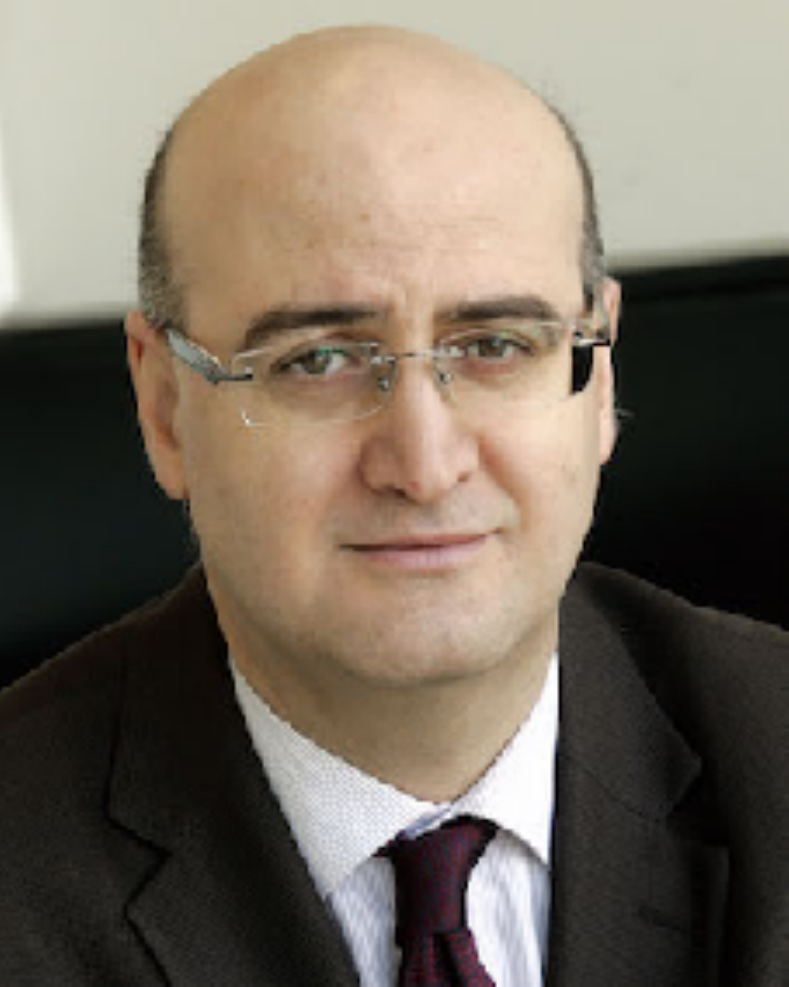}%
\end{minipage}}}]{Moncef Gabbouj}
 received his BS degree in electrical engineering in 1985 from Oklahoma
State University, and his MS and PhD degrees in electrical engineering
from Purdue University, in 1986 and 1989, respectively. Dr. Gabbouj
is currently Academy of Finland Professor and holds a permanent position
of Professor of Signal Processing at the Department of Signal Processing,
Tampere University of Technology. His research interests include multimedia
content-based analysis, indexing and retrieval, machine learning,
nonlinear signal and image processing and analysis, voice conversion,
and video processing and coding. Dr. Gabbouj is a Fellow of the IEEE
and member of the Finnish Academy of Science and Letters. He is the
past Chairman of the IEEE CAS TC on DSP and committee member of the
IEEE Fourier Award for Signal Processing. He served as Distinguished
Lecturer for the IEEE CASS. He served as associate editor and guest
editor of many IEEE, and international journals. 
\end{IEEEbiography}

\vfill{}

\end{document}